\documentstyle[12pt,doublespace]{article}
\begin{document}

\title{On the Geometry of Planar Domain Walls}
\author{
F. M. Paiva\thanks{Internet: fmpaiva@on.br} \ and Anzhong
Wang\thanks{Internet: wang@on.br} \\ \mbox{} \\ \small Departmento de
Astrof\'{\i}sica, Observat\'orio Nacional~--~CNPq, \\ \small Rua
General Jos\'e Cristino 77, 20921-400 Rio de Janeiro~--~RJ, Brazil}
\date{}

\maketitle

\begin{abstract}
The Geometry of planar domain walls is studied. It is argued that the
planar walls indeed have plane symmetry. In the Minkowski coordinates
the walls are mapped into revolution paraboloids.
\end{abstract}

{\sc pacs} numbers: 04.20.Jb, 04.20.Gz, 98.80.Cq, 11.27.+d.

\newpage

\section*{}

Domain walls are objects formed in the early stages of the evolution of
the Universe \cite{Kibble1976} and have been studied intensively in the
past decade or so, mainly because of their notable implications to
Cosmology \cite{Vilenkin1985}.  The first analytic solution of a domain
wall with plane symmetry\footnote{Here we use the definition for plane
symmetry originally given by Taub \cite{Taub1951}. That is, the
symmetric plane has three Killing vectors, two represent the
translation symmetry and one represents the rotation. Recently, this
definition was generalized to a more general case
\cite{Centrella1979,Wang1992a}.} was found by Vilenkin in 1983
\cite{Vilenkin1983}. Since then, this solution has been studied by
several authors, among them are Gibbons \cite{Gibbons1983} and Wang and
Letelier \cite{WangLetelier1995}, mainly concerning the global
structure of the spacetime. Despite of its simplicity, the solution
exhibits a very rich global structure. In particular, in each of the
three spatial directions there is a horizon. In the direction
perpendicular to the wall the horizon is not stable against the
perturbations of null fluids \cite{WangLetelier1995} and massless
scalar fields \cite{Wang1992b}. In 1984, on the other hand, Ipser and
Sikivie \cite{IpserSikivie1984} found all the planar domain wall
solutions that connect two flat regions.

In the existing literature, it is usually believed that those planar
domain walls do not really have plane symmetry but spherical one.
This belief is mainly due to the early studies of domain walls
\cite{IpserSikivie1984}. In fact, it was showed that
{\em ``In Minkowski coordinates, this planar domain wall is not a plane
at all, but rather an accelerated sphere}". This conclusion was also
reached in \cite{Ipser1984}: {\em ``In each case the wall is bent into
a closed surface enveloping the original } $z > 0$ {\em side of the
wall}", and used quite recently in \cite{KodamaIshiharaFujiwara1994} to
study the gravitational radiation of planar domain walls.

However, it is well-known that a
plane has no one-to-one mapping to a spherical surface.  So, it is very
curious to see how a planar domain wall is bent into a bubble. In Ref.
\cite{WangLetelier1995}, by considering the analytic maximum extension
of the spacetime, it was argued that the geometry of a planar domain
wall is a plane. In this short Comment, we shall stress the same
argument but in a different direction, and show that in the Minkowski
coordinates a planar domain wall is mapped into a revolution paraboloid,
instead of a spherically symmetric bubble.

Before proceeding, we would like first to clarify the following
concepts:  First, when we talk about the geometry of a wall, we mean
the geometry of the space-like two-surface of the wall. Second, the Nambu
action for a domain wall \cite{Vilenkin1985} is defined in a
(2+1)-dimensional hypersurface (or a tube), which represents the whole
history of the evolution of the wall. Bearing the above in mind, let us
consider the Vilenkin domain wall solution
\begin{equation}
ds^2 = e^{- k|z|} \{ dt^2 - dz^2 - e^{kt}(dx^{2} + dy^{2})\}, \label{ds2}
\end{equation}
where $k$ is a positive constant, and the range of the coordinates is
$ -\infty < t, z, x, y < + \infty$. The coordinates will be numbered as
$\{x^{\mu}\} = \{t, z, x, y\}, \; (\mu = 0, 1, 2, 3)$. The
corresponding energy-momentum tensor is given by $T^{\mu}_{\nu} =
2k\{1, 0, - 1, - 1\} \delta(z)$. Thus, it represents a planar domain
wall with support only on the hypersurface $z = 0$. The wall has the
plane symmetry characterized by the three Killing vectors
$\partial_{x}, \partial_{y},$ and $y\partial_{x} - x\partial_{y}$,
which act on the two-dimensional surfaces $t$, $z$ constant.

Using Cartan scalar techniques \cite{MacCallumSkea1994}, we were able
to show that besides these three Killing vectors the spacetime has
other three. Indeed, the non-zero Cartan scalars\footnote{
The Cartan scalars are basically the components of the Riemann tensor
and its covariant derivatives calculated in a constant frame
providing a complete local characterization of spacetimes
\cite{Cartan1951,Karlhede1980}. Spinor components are
used and the relevant objects here are the Ricci
spinor $\Phi_{AB'}$, the curvature scalar $\lambda$ and its first
symmetrized covariant derivatives $\nabla\Phi_{AB'}$ and
$\nabla\lambda_{AB'}$. {\sc sheep} and {\sc classi}
\cite{Frick1977,Aman1987} were used in the calculations.}
for the metric (\ref{ds2}) are
\begin{equation}
\begin{array}{rcl}
\Phi_{11'} = -2\Phi_{00'} = -2\Phi_{22'} = - \Lambda &=&
- \frac{1}{4}k^2\delta(z), \\
\nabla\Phi_{11'} = - \nabla\Phi_{22'} = 3\nabla\Phi_{33'} =
-3\nabla\Phi_{00'} &=& \\
-\frac{2}{3}\nabla\lambda_{00'} = \frac{2}{3} \nabla\lambda_{11'} &=&
- \frac{\sqrt{2}}{12}k^3 \delta'(z),
\end{array}
\end{equation}
where $\delta(z)$ denotes the Dirac delta function and $\delta'(z)$ its
derivative.  Outside the hypersurface $z = 0$, all Cartan scalars
vanish, therefore, in this region the spacetime is locally flat.
Otherwise, from this set of Cartan scalars, one finds that the metric
is Segre type [(1,11)1] and its isotropy group is SO(2,1), which is
three dimensional, leading to three Killing vectors (one of them is the
spatial rotation $y\partial_x - x\partial_y$ on the plane of
symmetry).  Since $z$ is the only coordinate appearing on the Cartan
scalars, $z = Const.$ are three dimensional homogeneous hypersurfaces,
leading to other three Killing vectors (two of them are the
translations $\partial_x$ and $\partial_y$ on the plane of symmetry).
Thus, the spacetime has six Killing vectors which act on the
$(2+1)$-hypersurfaces $z$ constant.  Three of them, $\partial_{x}$,
$\partial_{y}$ and $x\partial_{y} - y\partial_{x}$ act on the
2-dimensional surfaces $t$, $z$ constant.  Because these
$(2+1)$-hypersurfaces are time-like, a time-like Killing vector must
exist. Thus, the spacetime is at least stationary. In fact, it is
locally static, as can be seen by performing the following coordinate
transformations
\begin{equation}
\begin{array}{lcl}
t &=& \bar{t} + \alpha \ln\left[\frac{(\alpha^{2} -
      \rho^{2})^{1/2}}{\alpha^{2}} \right], \\
x &=& \frac{\alpha^{2} \rho e^{- \bar{t}/\alpha}}{(\alpha^{2} -
      \rho^{2})^{1/2}} \cos\phi, \\
y &=& \frac{\alpha^{2} \rho e^{- \bar{t}/\alpha}}{(\alpha^{2} -
      \rho^{2})^{1/2}} \sin \phi,
\end{array}
\end{equation}
where $\alpha \equiv 2/k$. In terms of $\bar{t}, \rho$ and $\phi$, the
metric (1) takes the form
\begin{equation}
ds^2 = e^{- k|z|} \{ f(\rho) d{\bar{t}}^2 - f^{-1}(\rho) d\rho^2 -
\rho^2 d\phi^{2} - dz^2\},
\end{equation}
where
\begin{equation}
f(\rho) = 1 - \frac{\rho^{2}}{\alpha^{2}},
\end{equation}
and $-\infty < \bar{t}, z < +\infty$, $ 0 \leq \rho < \alpha$ and $0 \leq
\phi \leq 2\pi$, which covers part of the spacetime (1). It should be
stressed that although the solution is static in the above domain,
globally it is not, as Gibbons pointed out in a different
way \cite{Gibbons1983}.

To study the geometry of the wall, following Ref.
\cite{IpserSikivie1984} (see also Ref. \cite{Gibbons1983}), let us make
the following coordinate transformations
\begin{equation}
\begin{array}{lcl}
T &=& \frac{e^{-kz/2}}{4k} \left\{ k^{2}(x^{2} + y^{2}) e^{-kz/2} +
      8 \sinh\left(\frac{kt}{2}\right) \right\},   \\
Z &=& \frac{e^{-kz/2}}{4k} \left\{ k^{2}(x^{2} + y^{2}) e^{-kz/2} -
8 \cosh\left(\frac{kt}{2}\right) \right\}, \\
X &=&  e^{k(t - z)/2}x, \hspace{1cm} Y =  e^{k(t - z)/2}y,
\end{array}
\end{equation}
or inversely
\begin{equation}
\begin{array}{lcllcl}
t &=& \frac{2}{k}\ln\left[\frac{T - Z}{(R^{2} - T^{2})^{1/2}}\right], &
z &=& \frac{2}{k}\ln\left[\frac{2}{k(R^{2} - T^{2})^{1/2}}\right], \\
x &=& \frac{2X}{k(T - Z)}, & y &=& \frac{2Y}{k(T - Z)},
\end{array}
\end{equation}
in the region $z \ge 0$, where $R^{2} \equiv X^{2} + Y^{2} + Z^{2}$.
Replacing $z$ by $- z$ in the above equations, we will get the
coordinate transformations in the region $z \le 0$.  Because of the
reflection symmetry, without loss of generality, we shall focus our
attention in the region $z \ge 0$.

In terms of $T, X, Y,$ and $Z,$ the metric (1) becomes
$ ds^{2} = dT^{2} - dX^{2} - dY^{2} - dZ^{2}$ in the region $z \ge 0 $,
which has the Minkowski form. Thus, we shall refer the coordinates $T,
X, Y,$ and $Z$ to as the Minkowski coordinates. From Eq.(6), on the
other hand, we find that the two-dimensional surfaces $t = Const.$, say,
$ t = t_{0}$, and $z = 0$ are given by
\begin{eqnarray}
X^{2} + Y^{2} + Z^{2} - T^{2} &=& \frac{4}{k^{2}}, \\
T - Z &=& \frac{2}{k}e^{t_{0}k/2}.
\end{eqnarray}
Thus, we have
\begin{equation}
X^{2} + Y^{2} =
\frac{4e^{t_{0}k/2}}{k}
\left\{\frac{2}{k}\cosh\left(\frac{t_{0}k}{2} \right) + Z \right\},
\end{equation}
which represents a revolution paraboloid. Therefore, the planar wall
looks like a revolution paraboloid in the Minkowski coordinates, instead
of a spherically symmetric bubble \cite{IpserSikivie1984,Ipser1984}.
Because of the reflection symmetry, we will have the same conclusion
when we are working in the region $z \le 0$.

On the other hand, that the planar wall is not a bubble can be also stressed
by considering the Killing vectors. Let us first assume that the wall is a
spherically symmetric bubble in the Minkowski coordinates. Then, we know
that in this coordinate system the three Killing vectors that define the
spherical symmetry are
\begin{eqnarray*}
\xi_{(1)} &=& Y\frac{\partial}{\partial X} - X\frac{\partial}{\partial Y}
        =  y\frac{\partial}{\partial x} - x\frac{\partial}{\partial y},\\
\xi_{(2)} &=&  Z\frac{\partial}{\partial Y} - Y\frac{\partial}{\partial Z}\\
        &=& \frac{e^{-kt}}{4k}\left\{4kye^{kt}\frac{\partial}{\partial t}
-2k^{2}xye^{kt}\frac{\partial}{\partial x} - \left[ k^{2}(y^2 - x^2)e^{kt}
+ 4(1 + e^{kt})\right]\frac{\partial}{\partial y}\right\},\\
\xi_{(3)} &=&  Z\frac{\partial}{\partial X} - X\frac{\partial}{\partial Z}\\
          &=& \frac{e^{-kt}}{4k}\left\{4kxe^{kt}\frac{\partial}{\partial t}
-2k^{2}xye^{kt}\frac{\partial}{\partial y} - \left[ k^{2}(x^2 - y^2)e^{kt}
+ 4(1 + e^{kt})\right]\frac{\partial}{\partial x}\right\}.
\end{eqnarray*}
{}From the above expressions it is clear that only the Killing vector
$\xi_{(1)}$ acts on the 2-dimensional space-like surface, $t = Const.$
and $z = 0$, of the  wall, and $\xi_{(2)}$ and $\xi_{(3)}$ act outside
this surface. This contradicts to our assumption, since if the wall has
spherically symmetry, then the above three Killing vectors should act on it.
Therefore, it is concluded that the geometry of the Vilenkin planar
domain wall is plane, on which the three Killing vectors $\partial_{x},
\partial_{y},$ and $y\partial_{x} - x\partial_{y}$ act.

All the solutions, which represent infinitely thin domain walls
connecting two flat regions, were given in Ref.
\cite{IpserSikivie1984}.  By a similar consideration, one can show that
all these walls have the same topological structure as the Vilenkin
wall. Moreover, this is also true for the Goetz domain wall with
non-zero thickness \cite{Goetz1990,WP1994}.

Recently, Cveti\v{c} and cor-workers \cite{Cvetic1993} carried out a
detailed study of the topology of domain walls, including the planar
ones. In particular, it was shown that the only geodesically complete
$(2+1)$-dimesional spacetime $z = Const.$ is the one whose
two-dimensional spatial surfaces, $ t, z = Const.$ have
positive curvature, which means that the geometry of the wall in that
case is a compact bubble. On the other hand, from the Vilenkin solution
(1) we can see that the hypersurfaces $z = Const.$ are the
$(2+1)$-dimensional de Sitter space written in a coordinate system in
which it is geodesically {\em incomplete}, and that , as a result,
the
Vilenkin planar domain wall looks like a revolution paraboloid in the
four-dimensional Minkowski space rather than a bubble.

In Ref. \cite{KodamaIshiharaFujiwara1994}, using the conclusions
obtained in Refs. \cite{IpserSikivie1984} and \cite{Ipser1984}, that
all planar domain walls are actually bubbles, the gravitational
radiation of a wall was studied. In particular, it was found that to
the first-order approximation the wall does not emit gravitational
waves. As we know, it is much easier for an object with plane symmetry
to emit gravitational waves than for one with spherical symmetry. So,
it would be very interesting to consider the problem directly in the
coordinates of Eq.(1). It is most likely that the situation will be
different and the final results will support the earlier speculations
of Vachaspati, Everett and Vilenkin \cite{VEV1984}.

\section*{Acknowledgment}
The authors gratefully acknowledge financial assistance from CNPq.

\end{document}